\documentclass[a4paper,12pt,openany]{article}
\usepackage[top=20truemm,bottom=20truemm,left=20truemm,right=20truemm]{geometry}
\usepackage[]{hyperref}
\usepackage{amsmath,colortbl,amssymb,amsthm,ascmac,mathrsfs,mathtools}
\usepackage[]{graphicx}
\graphicspath{{./figs/}}

\title{Probe strings on AdS accelerating black holes}
\author{Koichi Nagasaki}
\date{\today}

\begin{document}
\begin{center}
	{\LARGE Probe strings on AdS accelerating black holes}\\
\vspace{2cm}
	{\large Koichi Nagasaki}\footnote{koichi.nagasaki24@gmail.com}\\
\vspace{0.5cm}
	{\small Department of Physics, 
	Toho University,\\
	Address: 2 Chome-2-1 Miyama, Funabashi, Chiba 274-8510, Japan}
\end{center}
\vspace{1cm}

\abstract{
In this work we consider a spacial kind of spacetime called AdS accelerating black holes.
This is a kind of black holes which contain a stringlike singularity along polar axises attached to the black hole and it accelerates the black hole.
By using a string as a probe we study the properties of complexity growth of black holes following the CA duality.
Our result is the growth of complexity is independent of acceleration but the string probe detects the effects of acceleration.
}

\paragraph{Keywords:}
black holes, holographic complexity, accelerating black holes, AdS/CFT correspondence, conical deficit
\section{Introduction}
An interesting goal of quantum gravity and cosmology is to reveal the inside of the black hole horizon or information problem of black holes \cite{Coleman:1991ku, Preskill:1992tc, Giddings:1993vj, Russo:2005aw, Sekino:2008he, Terno:2009cc, Hartman:2013qma, Bradler:2013gqa, Polchinski:2016hrw, Marolf:2017jkr}.
In order to reveal such a mystery, complexity is getting attention in recent research \cite{Maldacena:2013xja, Stanford:2014jda, Susskind:2014moa, Susskind:2014rva, Barbon:2015ria, Barbon:2015soa, Couch:2016exn}.
Especially it is expected to relate with the transparency of black hole horizon, the existence of ``firewalls" and black hole information paradox \cite{Hayden:2007cs, Harlow:2013tf, Mann:2015luq, Zhao:2017iul}.
By the holographic principle \cite{Maldacena:1997re, Aharony:1999ti}, one can expect that the complexity on the boundary CFT is dual to the ``holographic complexity" defined in the bulk gravity theory. 
If this is true, the black hole has dual quantities and it is helpful to answer the questions about black hole physics stated above.
There are two reliable conjectures --- CV and CA.
The first one states that the holographic complexity is equal to the maximal volume of the Einstein Rosen bridge. 
It is called complexity - volume (CV) \cite{Susskind:2014rva, Stanford:2014jda}.
The other one states that the holographic complexity is equal to the action calculated in a certain space time region. 
It is called complexity - action (CA) \cite{Brown:2015bva, Brown:2015lvg}.
In this paper we focus on the CA conjecture.
According to this conjecture, the black hole complexity is
\begin{equation}
\mathcal C = \frac{S_\text{WDW}}{\pi\hbar},
\end{equation}
where $S_\text{WDW}$ is the action on the Wheeler-DeWitt patch.
The Lloyd bound is stated in \cite{2000Natur4061047L}.
CA conjecture is tested in various spacetime setting \cite{Pan:2016ecg, Momeni:2016ekm, Chapman:2016hwi, Lehner:2016vdi, Carmi:2016wjl, Tao:2017fsy, Alishahiha:2017hwg, Reynolds:2017lwq, Qaemmaqami:2017lzs, Guo:2017rul, Miao:2017quj, Sebastiani:2017rxr, Couch:2017yil, Swingle:2017zcd, Cano:2018aqi, Ghaffarnejad:2018bsd, Chapman:2018dem, Chapman:2018lsv, Fareghbal:2018ngr, Auzzi:2018pbc, Ghaffarnejad:2018prc, Alishahiha:2018tep, An:2018xhv}.

Here we add a probe string.
The whole system is described by the above Einstein Hilbert action and the Nambu-Goto action where the integration is performed over the Wheeler De-Witt patch, a specific region characterized by boundary time.
This string extends to AdS boundary and then the Lloyd bound is satisfied.

Accelerating black holes \cite{PLEBANSKI197698, Podolsky:2002nk} is an important kind of black holes. 
However, the complexity of these black holes is not well studied ever. 
The spacetime of such a black hole spacetime is described by a class of ``$\mathbf C$-metric" 
\cite{PhysRevD.2.1359, Bicak:1989knv, PhysRevD.67.064001, Bicak:1999sa, Griffiths:2005qp, Anabalon:2018qfv, Zhang:2019vpf}, and for AdS cases, see \cite{Podolsky:2002nk}.
The holographic complexity on these black holes are studies in \cite{Jiang:2021pzf,Chen:2021qbs}
In \cite{Chen:2021qbs} it is found that the growth rate is independent of the acceleration parameter while it depends only on the deficit angle $K$ for which $K=1$ case it recovers the usual AdS Schwarzschild black holes. 
According to them, the complexity growth at late time is
$d\mathcal C/dt = 2m/(\pi K)$.
For the range $0<K<1$ this growth rate does not respect the Lloyd bound while in the range $K\geq1$ this follows the Lloyd bound. 

To look the effects of acceleration we introduce a probe string in this work.
For flat spacetime case this method is applied to find a relation of drag forces and energy loss of the charged quarks \cite{Gubser:2006bz}.
Before we studied the effects of a probe sting on the AdS Schwarzschild black holes in \cite{Nagasaki:2017kqe,Nagasaki:2018csh,Nagasaki:2019icm}.
In the recent research this method is applied for the case of $(n+2)$-dimensional massive gravity \cite{Santos:2020lmb, Santos:2020xox, Zhou:2021vsm}.

\section{Accelerating black holes}\label{sec:acceleratingBH}
We first review AdS accelerating gravity in four dimension. 
The action of the Einstein gravity with a negative cosmological constant is
\begin{equation}
S_\text{bulk} = \frac1{16\pi}\int_{\mathcal M}d^4x\sqrt{-g}(R-2\Lambda),
\end{equation}
where $R$ is the scalar curvature and the cosmological constant is given by the AdS radius as 
$\Lambda = -6/\ell^2$.

The AdS accelerating black hole is known as a solution of this system. 
This solution is a kind of $\mathbf C$-metric \cite{Griffiths:2005qp}.
That metric is given \cite{Hong:2003gx} by
\begin{equation}
ds^2 = \frac1{\Omega^2}\bigg[-f(r)dt^2 + \frac{dr^2}{f(r)} 
 + r^2\Big(\frac{d\theta^2}{\Sigma} + \Sigma\sin^2\theta\frac{d\phi^2}{K^2}\Big)\bigg].
\end{equation}
In the above the functions are defined as
\begin{align}
f(r) &= (1-\alpha^2r^2)\Big(1 - \frac{2m}{r}\Big) + \frac{r^2}{\ell^2},\\
\Omega &= 1 + \alpha r\cos\theta,\quad
\Sigma = 1 + 2m\alpha\cos\theta.
\end{align}

The physical singularity is located at $r=0$ and the horizon is determined by $f(r_h)=0$.
Here $\Omega$ is the conformal factor which determines the conformal boundary of the spacetime.
$\alpha$ is the acceleration parameter.
In the following discussion, we use the length unit where AdS radius is $\ell = 1$ for simplicity.
The acceleration horizon is the point where 
$f(r_\text{ah})|_{m=0} = 0$.
 Explicitly this is 
$f_0(r_\text{ah}) = 0$ that is
$1 + (\ell^{-2} - \alpha^2)r_\text{ah}^2 = 0$,
$r_\text{ah} = \ell/\sqrt{\alpha^2\ell^2-1}$.

\paragraph{Special cases}
The above solution includes the following parameters --- the scale of the acceleration $\alpha$, the deficit angle $K$, AdS radius $\ell$ and the black hole mass $m$.
For $m=0$ case the coordinate transformation 
\begin{equation}
1 + \frac{R^2}{\ell^2} = \frac{f(r)}{(1-A^2\ell^2)\Omega^2},\qquad
R\sin\Theta = \frac{r\sin\theta}{\Omega}
\end{equation}
gives a pure AdS spacetime metric \cite{Podolsky:2002nk}.

$K$ gives the deficit angle around the axis
$\delta = 2\pi(1-1/K) = 8\pi\mu$, where $\mu$ is the tension of the non-singular cosmic string which gives the acceleration to the black hole \cite{Gregory:2017ogk,Gregory:2019dtq}. 
However, this can be eliminated from by setting $K_\pm = 1\pm2m\alpha$ so that the deficit angle at the north/south pole vanish \cite{Xu:2017nut}.

\section{Probe string effect}
Here we introduce a probe string.
In this case, it is still natural to assume that the string moves along the great circle of sphere part.
That motion is parametrized as \cite{Gubser:2006bz, Nagasaki:2017kqe, Nagasaki:2018csh, Nagasaki:2019icm, Zhou:2021vsm}:
\begin{equation}\label{eq:string_parametrization}
t = \tau,\;\;
r = \sigma,\;\;
\phi = \omega\tau + \xi(\sigma).
\end{equation}
In this setting we calculate the Nambu-Goto (NG) action:
\begin{equation}
S_\text{NG} = \int d\tau d\sigma\mathcal L,\;\;
\mathcal L = \sqrt{-\det g_\text{ind}}.
\end{equation}
Here the induced metric is
\begin{equation}
g_\text{ind}
= \begin{pmatrix}
\displaystyle{-\Big(f(\sigma) - \frac{\omega^2\sigma^2}{K^2}\Big)}& 
 \displaystyle\frac{\omega\sigma^2\xi'}{K^2}\\
\displaystyle\frac{\omega\sigma^2\xi'}{K^2}& 
 \displaystyle\frac1{f(\sigma)} + \frac{\sigma^2\xi'^2}{K^2}
\end{pmatrix}.
\end{equation}

The unknown function in \eqref{eq:string_parametrization} is $\xi(\sigma)$. 
First we need to find this function. 
The equation of motion for $\xi$ is 
\begin{equation}
\frac{\sigma^2f\xi'}{K^2}\Big(1 - \frac1{K^2}\frac{\omega^2\sigma^2}{f}
 + \frac{\sigma^2f}{K^2}\xi'^2\Big)^{-1/2} = c,
\end{equation}
where $c$ is the integration constant we determine later.
By solving the above equation for $\xi'(\sigma)$,
\begin{equation}
\xi'(\sigma) 
= \frac{cK^2}{|\sigma^2f|}\sqrt\frac{\sigma^2f - \omega^2\sigma^4/K^2}{\sigma^2f - c^2K^2},
\end{equation}
where we impose that $c$ is a real positive number and use the fact that $K$ is positive real valued: $0< K<\infty$.
We impose the reality condition to $\xi'(\sigma)$. 
It requires that the numerator and the denominator in the square root become zero at the same point.

We define $\sigma_H$ as $F(\sigma_H) - \omega^2\sigma_H^4/K^2 = 0$ where
$F(\sigma)\coloneqq\sigma^2f(\sigma)$.
The value of $\sigma_H$ is larger than $\geq r_h$ (horizon), {\it i.e.}, it is located outside of the horizon.
By the value $\sigma_H$, $c$ is determined as 
\begin{equation}
c^2 = \frac{F(\sigma_H)}{K^2}
= \frac{\omega^2\sigma_H^4}{K^4}.
\end{equation}
Then
\begin{equation}
\xi'(\sigma) = \frac{\omega\sigma_H^2}{|F|}\sqrt\frac{F - \omega^2\sigma^4/K^2}{F - \omega^2\sigma_H^4/K^2}.
\end{equation}

The numerator in the square root is 
\begin{equation}
F(\sigma) - \frac{\omega^2}{K^2}\sigma^4
= (1-\alpha^2\sigma^2)(\sigma^2-2m\sigma) + \Big(1 - \frac{\omega^2}{K^2}\Big)\sigma^4.
\end{equation}

Since the denominator 
\begin{equation}
F(x) - c^2K^2 
= (1-\alpha^2)x^4 + 2m\alpha^2x^3 + x^2 - 2mx - c^2K^2,\quad
(0\leq \alpha < 1)
\end{equation}
goes to infinity when $\sigma\rightarrow\infty$, $F(\sigma) - (\omega^2/K^2)\sigma^4$ must behave in the same way in order to maintain the reality. 
This imposes $1-\alpha^2-\omega^2/K^2 > 0$.
\begin{equation}\label{eq:stringvel_limit}
|\omega| < K\sqrt{1-\alpha^2}.
\end{equation}
Then there is the upper limit for the string velocity.

\begin{figure}[h]
	\begin{minipage}[h]{0.5\linewidth}
	\begin{center}
	\includegraphics[width=9cm]{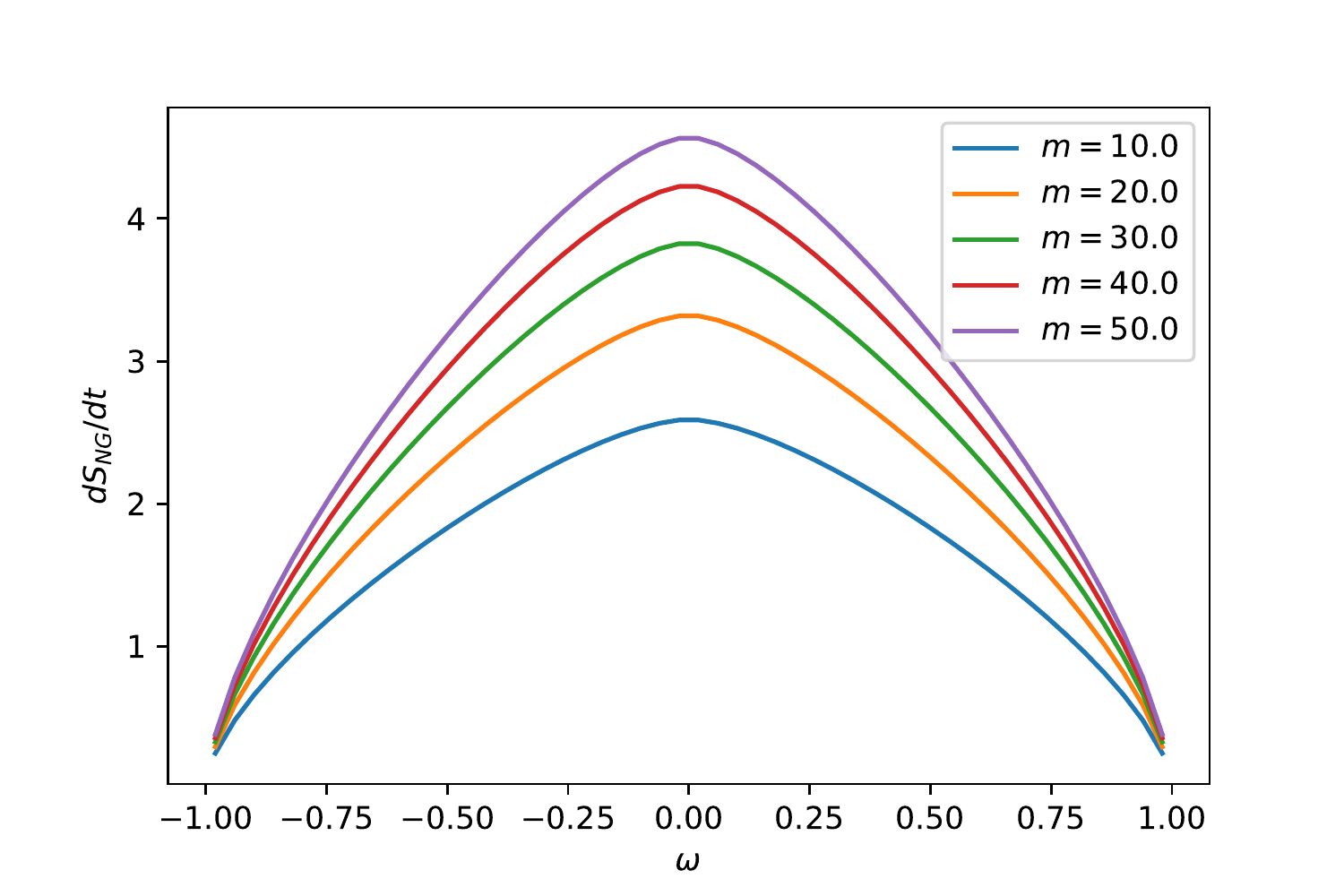}
	\caption{action - string velocity ($K=1$, $\alpha=0$)}
	\label{fig:NGaction_stringvelocity_K1_A0}
	\end{center}
	\end{minipage}
\hspace{0.01\linewidth}
	\begin{minipage}[h]{0.5\linewidth}
	\begin{center}
	\includegraphics[width=9cm]{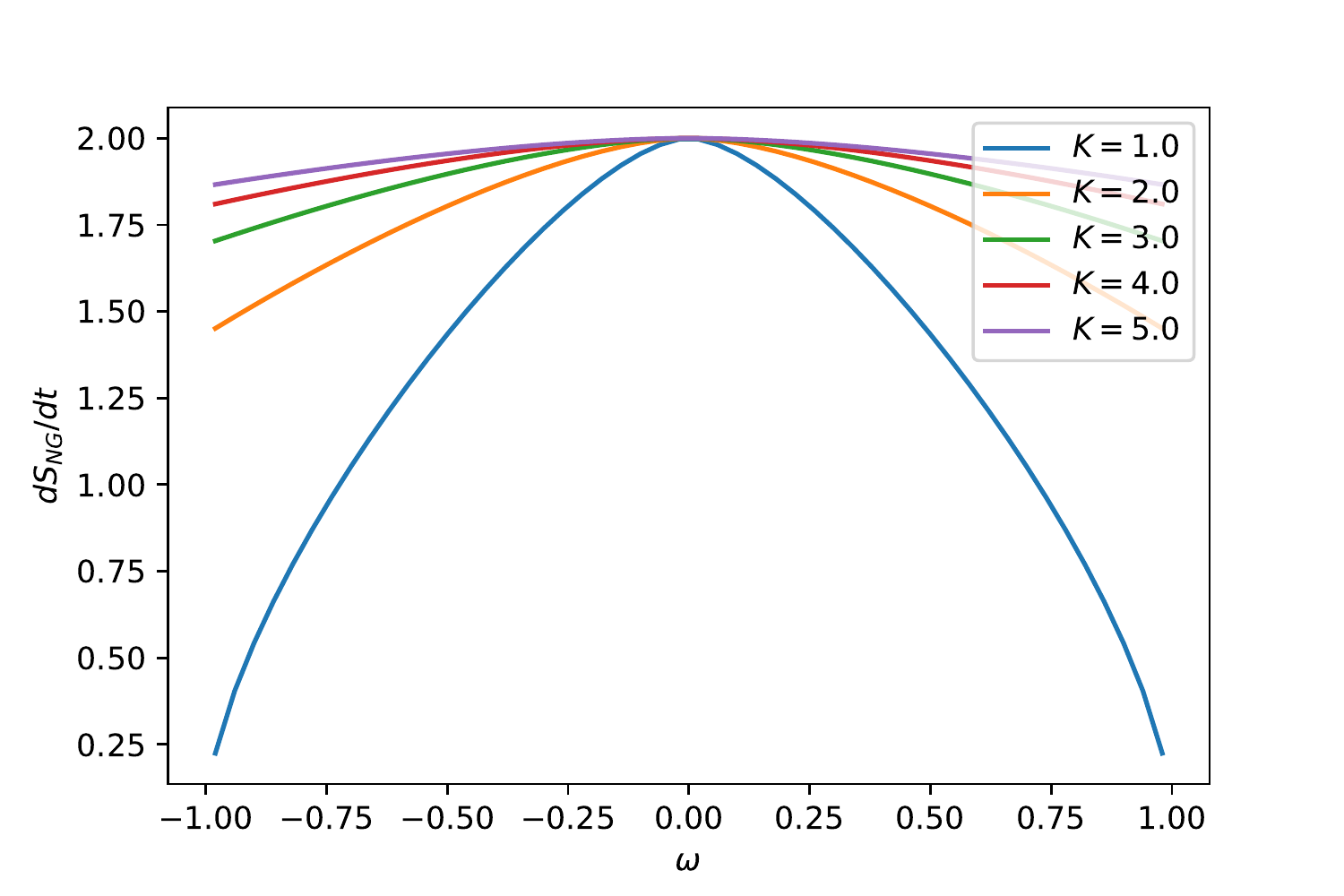}
	\caption{action - string velocity ($K>1$, $\alpha=0$, $m=5$)}
	\label{fig:NGaction_stringvelocity_Kmore1_A0}
	\end{center}
	\end{minipage}
\end{figure}

By summering them, the NG action is
\begin{align}
\frac{dS_\text{NG}}{dt}
&= \int d\sigma\sqrt{1 - \frac{\omega^2\sigma^4}{K^2}\frac1F + \frac{F}{K^2}\cdot\xi'^2}
= \int d\sigma\sqrt{\frac{K^2 - \omega^2\sigma^4/F}{K^2 - \omega^2\sigma_H^4/F}}.
\end{align}
Let us see the behavior of some solutions.
\paragraph{No deficit angle case ($K=1$)}
In this case, the action becomes
\begin{equation}
\frac{dS_\text{NG}}{dt}
= \int_0^{r_h}d\sigma\sqrt{\frac{\sigma^2f - \omega^2\sigma^4}{\sigma^2f - \omega^2\sigma_H^4}},
\end{equation}
where $\sigma_H$ is determined by $f(\sigma_H) - \omega^2\sigma_H^2 =0$.
This solution is plotted in Figure \ref{fig:NGaction_stringvelocity_K1_A0}. 
This reproduces the result of usual Schwarzschild black holes \cite{Nagasaki:2017kqe}.

\begin{figure}
	\begin{minipage}[h]{0.5\linewidth}
	\begin{center}
	\includegraphics[width=9cm]{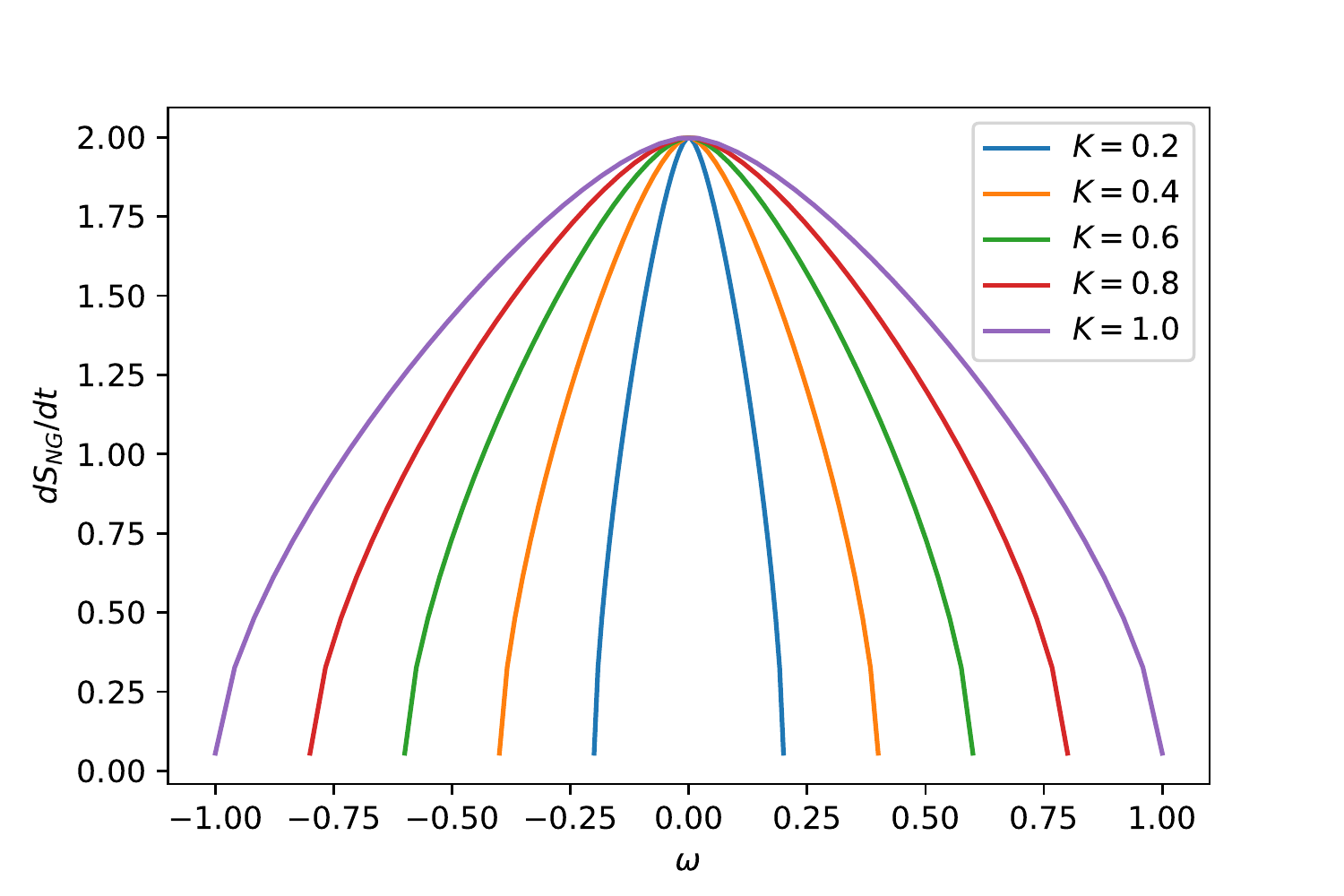}
	\caption{action - string velocity ($K<1$, $\alpha=0$, $m=5$)}
	\label{fig:NGaction_stringvelocity_Kless1_A0}
	\end{center}
	\end{minipage}
\hspace{0.01\linewidth}
	\begin{minipage}[h]{0.5\linewidth}
	\begin{center}
	\includegraphics[width=9cm]{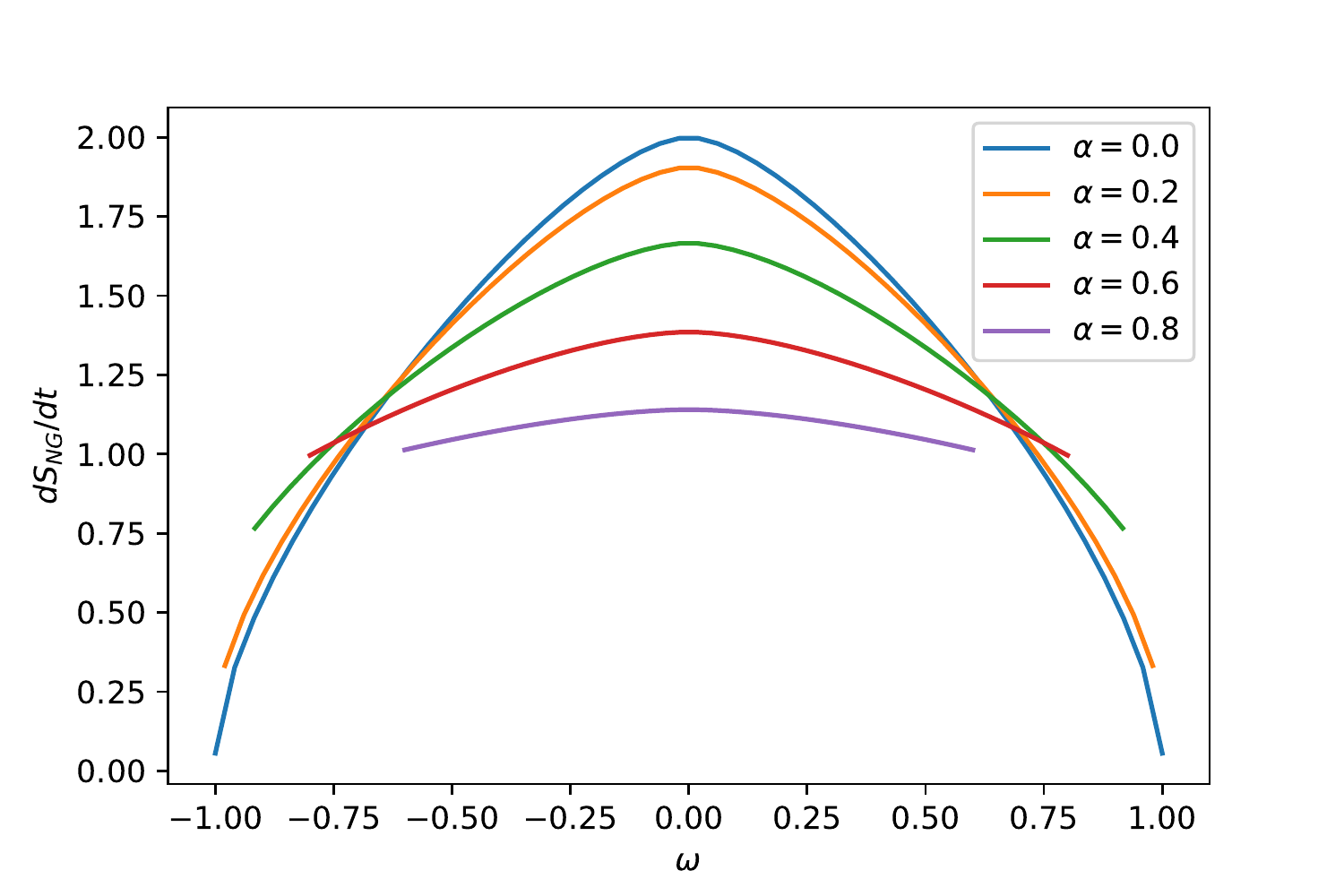}
	\caption{action - string velocity ($K=1$, $\alpha\geq0$, $m=5$)}
	\label{fig:NGaction_stringvelocity_K1_A}
	\end{center}
	\end{minipage}
\end{figure}

\paragraph{Non-trivial deficit angle ($K\neq1$)}
These cases are shown in Figure \ref{fig:NGaction_stringvelocity_Kmore1_A0} and Figure \ref{fig:NGaction_stringvelocity_Kless1_A0}.

\paragraph{Acceleration dependence}
The behavior for different acceleration parameter is shown in Figure \ref{fig:NGaction_stringvelocity_K1_A}.

As stated in Section \ref{sec:acceleratingBH}, the acceleration is thought to be caused by the cosmic string attached the black hole at the north and the south poles. 
The tension at the north ($+$) and south ($-$) poles are 
\begin{equation}\label{eq:tension_defangle}
8\pi\mu_\pm = 1 - \frac1K(1\pm 2m\alpha).
\end{equation}
Conventionally, we can set $K = 1+2m\alpha$ to set the deficit angle on the north pole to zero.
It makes the north pole regular while leaves the deficit angle on the south pole.
We have imposed the condition $1-\alpha^2-\omega^2/K^2 > 0$ on the string velocity \eqref{eq:stringvel_limit}.
Now this translates into the condition by the acceleration and mass:
\begin{equation}\label{eq:string_vel_limit}
|\omega| < (1+2m\alpha)\sqrt{1 - \alpha^2}.
\end{equation}

\begin{figure}
	\begin{minipage}[h]{0.5\linewidth}
	\begin{center}
	\includegraphics[width=9cm]{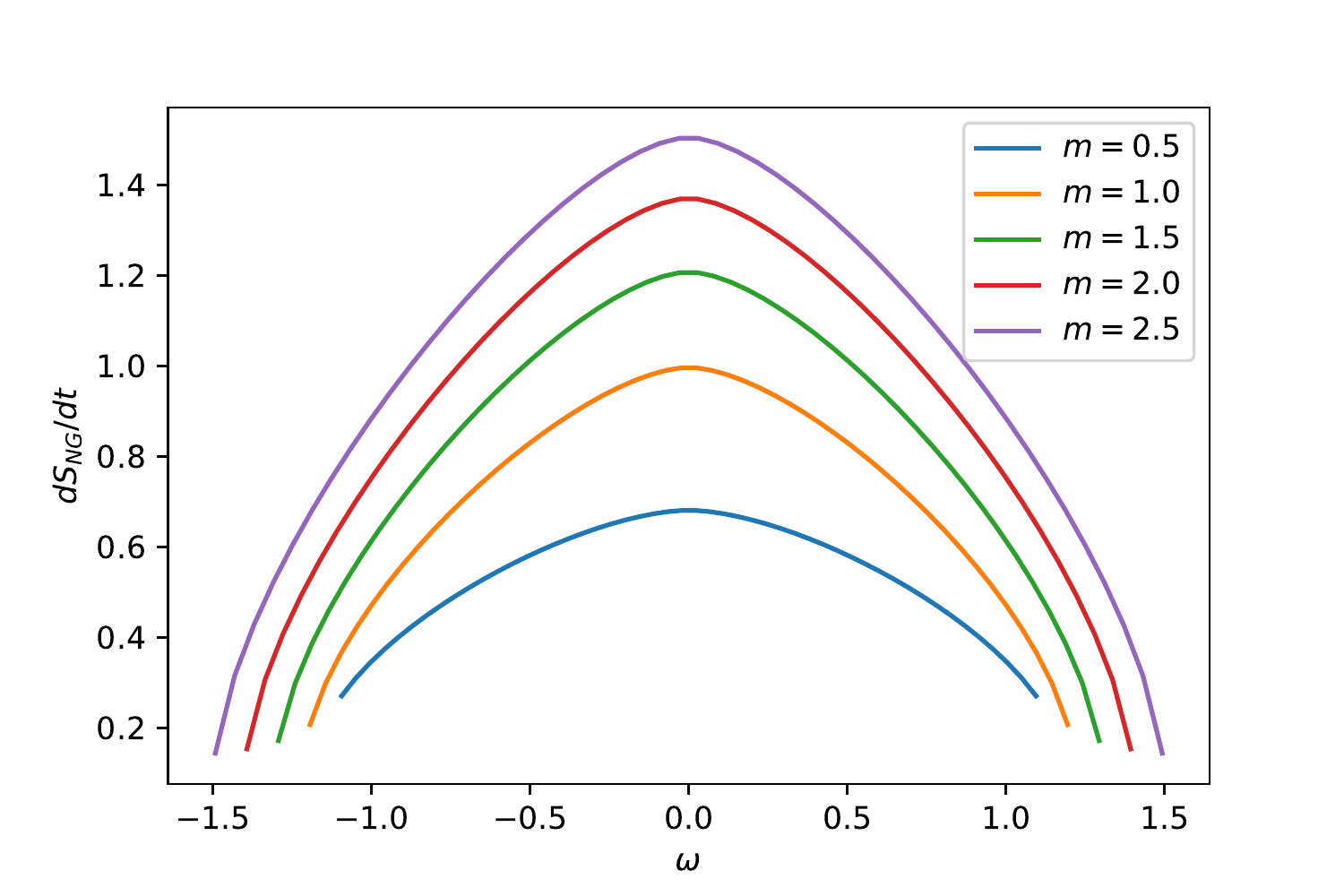}
	\caption{action - string velocity (different masses fixed $K$, $\alpha = 0.1$)}
	\label{fig:NGaction_stringvelocity_mass_Kfix}
	\end{center}
	\end{minipage}
\hspace{0.01\linewidth}
	\begin{minipage}[h]{0.5\linewidth}
	\begin{center}
	\includegraphics[width=9cm]{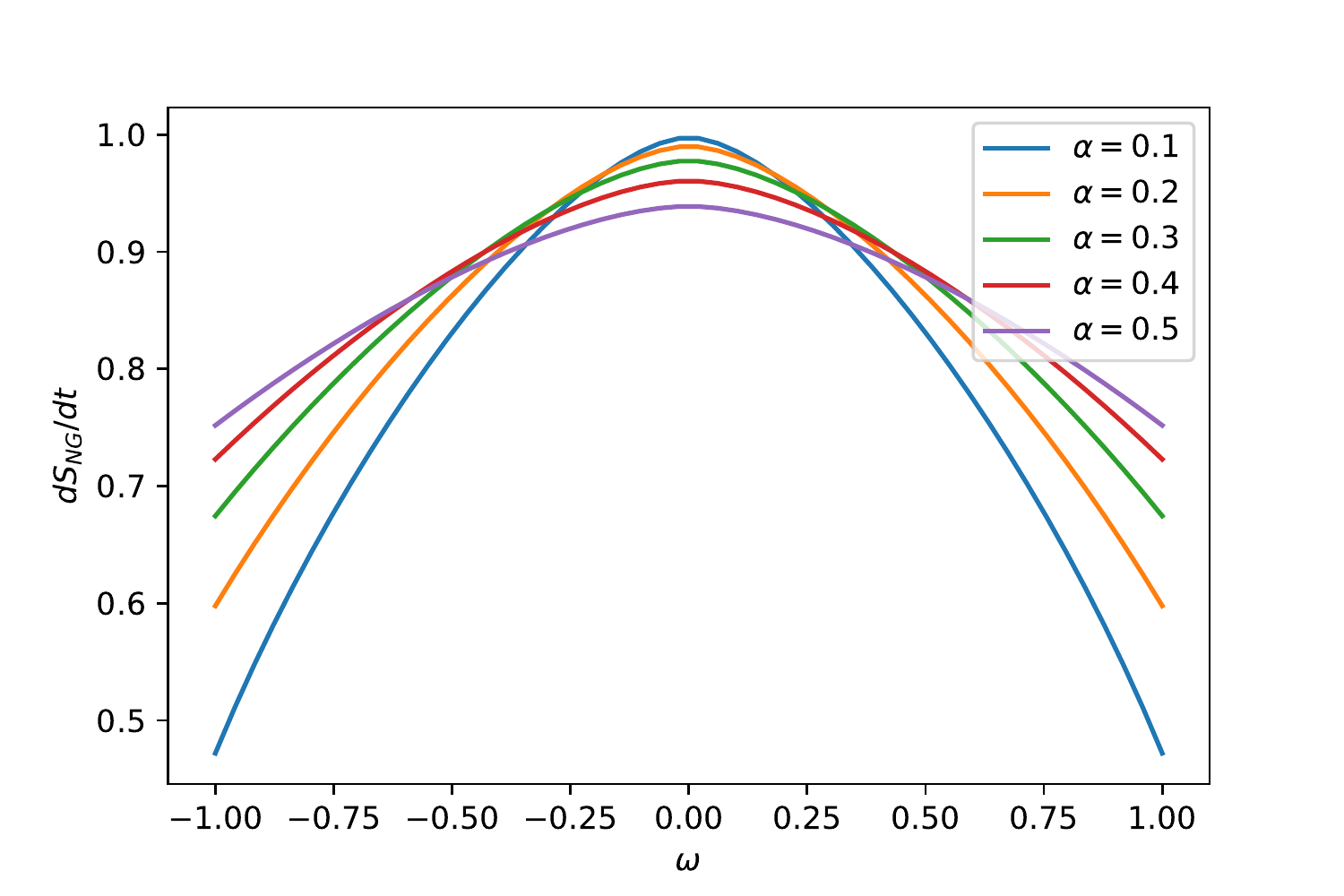}
	\caption{action - string velocity (different $\alpha$, $m=10$)}
	\label{fig:NGaction_stringvelocity_A_Kfix}
	\end{center}
	\end{minipage}
\end{figure}

By this convention the free parameters in this system are only the acceleration $\alpha$ and black hole mass.
Let us plot this dependence again.
The first figure (Fig.\ref{fig:NGaction_stringvelocity_mass_Kfix}) is the plot for string velocity dependence for different masses.
It says the effect to the complexity is an increasing function of mass in the same way as the usual Schwarzchild black holes. 
The next figure (Fig.\ref{fig:NGaction_stringvelocity_A_Kfix}) is the plot for different acceleration values.
It shows that the maximum value of the complexity growth decreases when the acceleration grows.
We can also see that even when the string velocity is zero, the complexity growth does not become zero.
It can be said that the acceleration decreases the effect to the growth of complexity.

\section{Discussion}
In this paper we studied the effect of the prove string on the AdS accelerating black hole.
The study of these kind of black holes is still unknown, especially in the perspective of CA conjecture.
Our new methods is using a prove string on these spacetime.
We summarize the new result in this study.

First, according to the result in \cite{Chen:2021qbs}, through the CA conjecture, the growth rate of the complexity is independent of the acceleration and only depends on the deficit angle $K$.
However, by using a rotating string as a probe we could find the acceleration dependence directly.
Notably, under the relation and the convention stated Eq.\eqref{eq:tension_defangle} and its below, the acceleration parameter $\alpha$ and the deficit angle $K$ are related and we can only focus on the acceleration parameter. 
That results are shown in the last two figures (Figure \ref{fig:NGaction_stringvelocity_mass_Kfix} and Figure \ref{fig:NGaction_stringvelocity_A_Kfix}).
Although the Lloyd bound is destroyed in the parameter range $0 < K < 1$, as stated in \cite{Chen:2021qbs}, under the convention \eqref{eq:tension_defangle}, $K$ does not take such values.   

Second, we also find that there is a limit for string velocities \eqref{eq:string_vel_limit}.
As we found before. the slower the probe string moved, the larger effect was given to the complexity growth \cite{Nagasaki:2017kqe}, 
Specifically its effect became zero at the vicinity of light speed.
For the accelerating case, there is a limit for the range of string velocities.
As shown in Figure \ref{fig:NGaction_stringvelocity_Kless1_A0} or Figure \ref{fig:NGaction_stringvelocity_mass_Kfix}, the effect of the string becomes zero at the upper limit of velocity.
Actually the plots do not make sense outside of the range $|\omega|<1$. 
In Figure \ref{fig:NGaction_stringvelocity_A_Kfix} we restrict the plot in the range $|\omega|<1$.
We conclude that the non-zero string effect survives at the light speed, unlike the usual Schwarzschild case.

Finally, there are other predictions for holographic complexity. 
One is the revised version of CV duality. 
It is called ``CV-2 " \cite{Couch:2016exn}.
The other is the revised version of CA duality. 
It is called ``CA-2" \cite{Fan:2018wnv}.
It is interesting to consider the effects of probe sting on these conjectures. 
Moreover, this method can determine which conjecture states the properties of holographic complexity.

\section*{Acknnowledgments}
This research is supported by Department of Physics, Toho University.

\providecommand{\href}[2]{#2}\begingroup\raggedright\endgroup

\end{document}